\def\h{${{h}_{75}}^{-1}$}
\def\lesssim{\mathrel{\hbox{\rlap{\hbox{\lower4pt\hbox{$\sim$}}}\hbox{$<$}}}}
\def\gtrsim{\mathrel{\hbox{\rlap{\hbox{\lower4pt\hbox{$\sim$}}}\hbox{$>$}}}}
\def\la{\mathrel{\hbox{\rlap{\hbox{\lower4pt\hbox{$\sim$}}}\hbox{$<$}}}}
\def\ga{\mathrel{\hbox{\rlap{\hbox{\lower4pt\hbox{$\sim$}}}\hbox{$>$}}}}

\documentclass{aa}
\usepackage{graphicx}

\def\cha{{\it Chandra\/}\ }
                       
\begin{document}

\title {\cha observation of the multiple merger cluster Abell 521}
\author{Ferrari, C.\inst{1}
\and Arnaud, M.\inst{2}
\and Ettori, S.\inst{3} 
\and Maurogordato, S.\inst{4} 
\and Rho, J.\inst{5} 
}
\offprints{Chiara Ferrari}  
\institute{
Institut f\"ur Astrophysik, Technikerstra{\ss}e 25, 6020 Innsbruck, Austria   
\and CEA$/$DSM$/$DAPNIA Service d'Astrophysique, CEA Saclay, L'Orme des
Merisiers, B\^at. 709, 91191 Gif-sur-Yvette, France
\and Osservatorio Astronomico di Bologna, Istituto Nazionale di Astrofisica, 
via Ranzani 1, I-40127 Bologna, Italy 
\and Laboratoire Cassiop\'ee, CNRS/UMR 6202, Observatoire de la C\^ote d'Azur, 
BP 4229, 06304 Nice Cedex 4, France
\and Spitzer Science Center, California Institute of Technology, Pasadena, CA
911251}

\date{Received~1 August 2005; accepted~8 September 2005}  

\abstract{
We present the \cha analysis of the rich galaxy cluster Abell~521
($z$=0.247). The high resolution of the \cha observation has allowed
us to refine the original merging scenario proposed for A521, and to
reveal new features in its X-ray emission. A521 has strongly
substructured ICM density and temperature maps. Its X-ray diffuse
emission is elongated along a NW/SE direction ({\it SX2}) and shows
two major components, a main cluster and a northern group of galaxies.
This latter is in turn substructured, showing a clump of cold and very
dense gas centred on the Brightest Cluster Galaxy (BCG), and a
northern tail aligned in the {\it SX2} direction. A compression of the
X-ray isophotes is also observed South of the BCG. We conclude that
the northern group is infalling onto the main cluster along the NW/SE
direction.  This hypothesis is corroborated by the presence of a hot
bar in the ICM temperature map located between the southern and
northern regions, as the gas could be compressionally heated due to
the subclusters' collision. The hot region corresponds to the eastern
part of an over-dense ridge of galaxies, along which it was originally
suggested that a merging of subclusters has recently occurred along
the line of sight. An alternative hypothesis about the origin of the
hot central bar is that we could observe in projection the shock
fronts due to this older cluster-cluster collision. However, the two
hypothesis do not exclude each other. Two other structures possibly
interacting with the main cluster are detected on the West and
North-East sides of the BCG. We also uncover the presence of two
northern edges in the ICM density, which could be due to the ongoing
merging events observed in the central field of the cluster, or even
in its outer regions. A521 is a spectacular example of a multiple
merger cluster made up by several substructures converging at
different epochs towards the centre of the system. The very pertubed
dynamical state of this cluster is also confirmed by our discovery of
a radio relic in its South-East region.
 }
\titlerunning{\cha observations of Abell 521}
\maketitle   

\section{Introduction}

In the concordant cosmological model ($\Lambda$CDM, $\Omega_{m}$=0.3
and $\Omega_{\Lambda}$=0.7), small structures are the first to form,
and then they merge giving rise to more and more massive systems in a
hierarchical way. Both numerical and observational results show that
galaxy clusters form and evolve through the merging of sub-clusters
and groups of galaxies along filamentary structures (e.g. West et
al. 1995; Bertschinger 1998; Durrett et al. 1998; Arnaud et al. 2000;
Bardelli et al. 2000; Borgani et al. 2004; Adami et al. 2005).

Combined optical and X-ray studies have been particularly successful
in revealing the dynamics of merging clusters (Flores et al. 2000;
Henriksen et al. 2000; Donnelly et al. 2001; Bardelli et al. 2002;
Barrena et al. 2002; Czoske et al. 2002; Rose et al. 2002; Valtchanov
et al. 2002; Boschin et al. 2004; Belsole et al. 2005; Demarco et
al. 2005; Durret et al. 2005; Ferrari et al. 2005). This field is more
and more active since precise spectro-imaging data in X-rays are now
available with \cha and XMM, allowing to derive high resolution
temperature and density maps, in which very typical signatures of
merging events have been detected such as strong temperature and
density variations (Markevitch \& Vikhlinin 2001; Belsole et al. 2004,
2005; Henry et al. 2004; Durrett et al. 2005), bow shocks (Markevitch
et al. 2002; Markevitch et al. 2005) and cold fronts (Markevitch et
al. 2000; Vikhlinin et al. 2001; Mazzotta et al. 2001; Sun et
al. 2002; Dupke et al. 2003). Detailed multi-wavelength studies of
galaxy clusters are essential to determine the scenario of their
formation, and to analyse the complex physical processes acting during
their evolution.
 
Abell 521 (z=0.247) is a relatively rich (R=1) cluster of Bautz-Morgan
Type III (Abell 1958; Abell et al. 1989). After its first detection in
X-ray with HEAO1 (Johnson et al. 1983; Kowalski et al. 1984), the
dynamical state of A521 has been investigated in detail through a
combined X-ray and optical analysis (Arnaud et al. 2000; Maurogordato
et al. 2000; Ferrari et al. 2003). A severe segregation between the
gas and galaxy distributions was detected. ROSAT/HRI observations
revealed the presence of two peaks of X-ray emission, associated with
a diffuse main cluster and with a compact less massive group (Arnaud
et al. 2000). Unlike what is usually observed in relaxed systems, the
Brightest Cluster Galaxy (BCG) is not located at the barycentre of
A521, but in the compact sub-group, with a surprising off-set from its
X-ray peak. The galaxy isodensity map in the central 20'${\times}$20'
field of A521 has a very irregular and strongly sub-clustered
morphology. Its general structure follows a NW/SE direction, crossed
by a perpendicular high density ridge of galaxies in the core region
(Arnaud et al. 2000; Ferrari et al. 2003). The analysis of the
dynamical and kinematic properties of more than one hundred cluster
members confirmed that A521 is far from dynamical equilibrium: its
radial velocity distribution significantly different from a Gaussian
and characterised by a very high dispersion ($1325 ^{+145}_{-100}$
km/s) is typical of merging systems (Ferrari et al. 2003). A detailed
dynamical analysis revealed at least two different and not
contemporary episodes of merging: a) a dynamically bound complex of
galaxies, hosting the BCG and corresponding to the compact group
detected in X-ray, is currently infalling on the plane of the sky
toward the centre of the main cluster, and b) two or more sub-clusters
have recently collided along the over-dense central {\it ridge}, with
a collision axis nearly along the line of sight (Ferrari et al. 2003).

The recent analysis by Umeda et al. (2004) of A521 ${\rm H}_{\alpha}$
luminosity function showed that this cluster contains more currently
star-forming galaxies than local clusters, consistently with the
observed Butcher-Oemler effect. The excess of star formation (SF) can
be at least partly related to the particular dynamical state of A521,
since an increase of SF has been observed in several merging systems
(e.g. Gavazzi et al. 2003; Poggianti et al. 2004; Ferrari et
al. 2005).

The complex dynamical state of A521 and its unique morphological
features motivated our \cha observations with the aim of better
characterising the physics of this exceptional cluster. In this paper
the \cha data are analysed. Sect.~\ref{ObsData} briefly describes the
observations and the data reduction. In Sect.~\ref{morphology} we
study the X-ray morphology and the temperature structure of
A521. Results are discussed in Sect.~\ref{disc} and summarised in
Sect.~\ref{summ}.  As in Ferrari et al. (2003), all numbers are
expressed as a function of $h_{75}$, the Hubble constant in units of
75 km ${\rm s^{-1}~Mpc^{-1}}$. We have used the $\Lambda$CDM model
with $\Omega_m=0.3$ and $\Omega_{\Lambda}=0.7$, thus 1~arcmin
corresponds to $\sim$0.217~\h~Mpc in the following.

\section{Observations and data reduction}\label{ObsData}

A521 was observed with \cha ACIS-I and ACIS-S in "VFAINT" mode.  The
datasets were processed and cleaned using CIAO 3.2 software and
calibration files in CALDB 3.0.0.  The first exposure was done on Dec
23, 1999 with ACIS-I and focal plane temperature of $-110^o$ for an
effective exposure time of 38.0 ksec after standard cleaning (88\% of
the nominal exposure time). On Oct 13, 2000, a second exposure of 41
ksec was done with ACIS-S and focal plane temperature of $-120^o$.
After cleaning the light curve from the several flares present by
requiring a mean count rate of 0.085 cts/s, an exposure of 18.4 ksec
is obtained.

\section{Results}\label{morphology}

\subsection{X-ray morphology}

\begin{figure*} 
\centering
\resizebox{18cm}{!}{\includegraphics{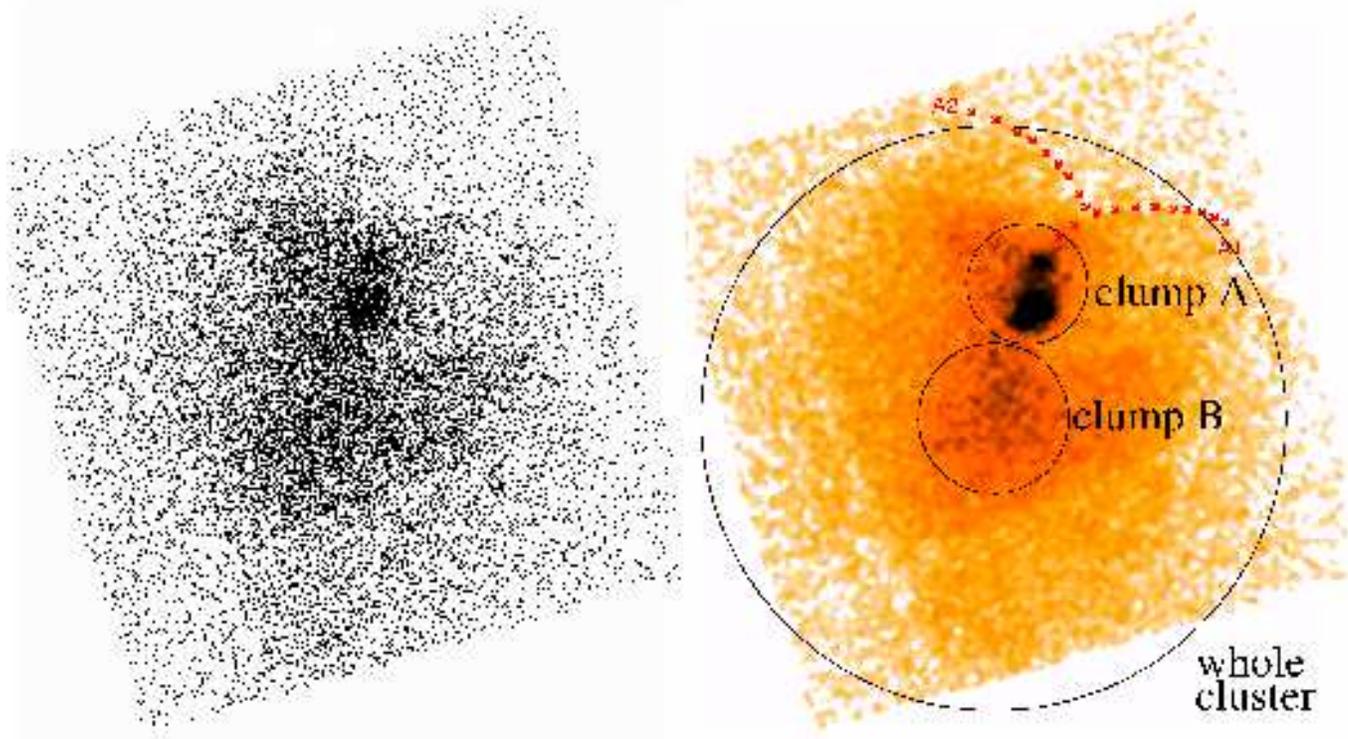}}
\hfill
\parbox[b]{18cm}{
\caption{0.4-5 keV raw image of A521 diffuse emission
(ACIS-S3 field) derived by summing the ACIS-I and ACIS-S
observations. In the right panel the image has been smoothed with a
Gaussian of $\sigma$=2 arcsec, and two possible arc-like
discontinuities are indicated in red. The ACIS-I CCD gaps are evident
in the East and South sides. The whole cluster region and its two main
clumps {\it A} and {\it B} are also indicated.}
\label{strucKT}}
\end{figure*}

\begin{figure*}
\centering
\resizebox{14cm}{!} {\includegraphics{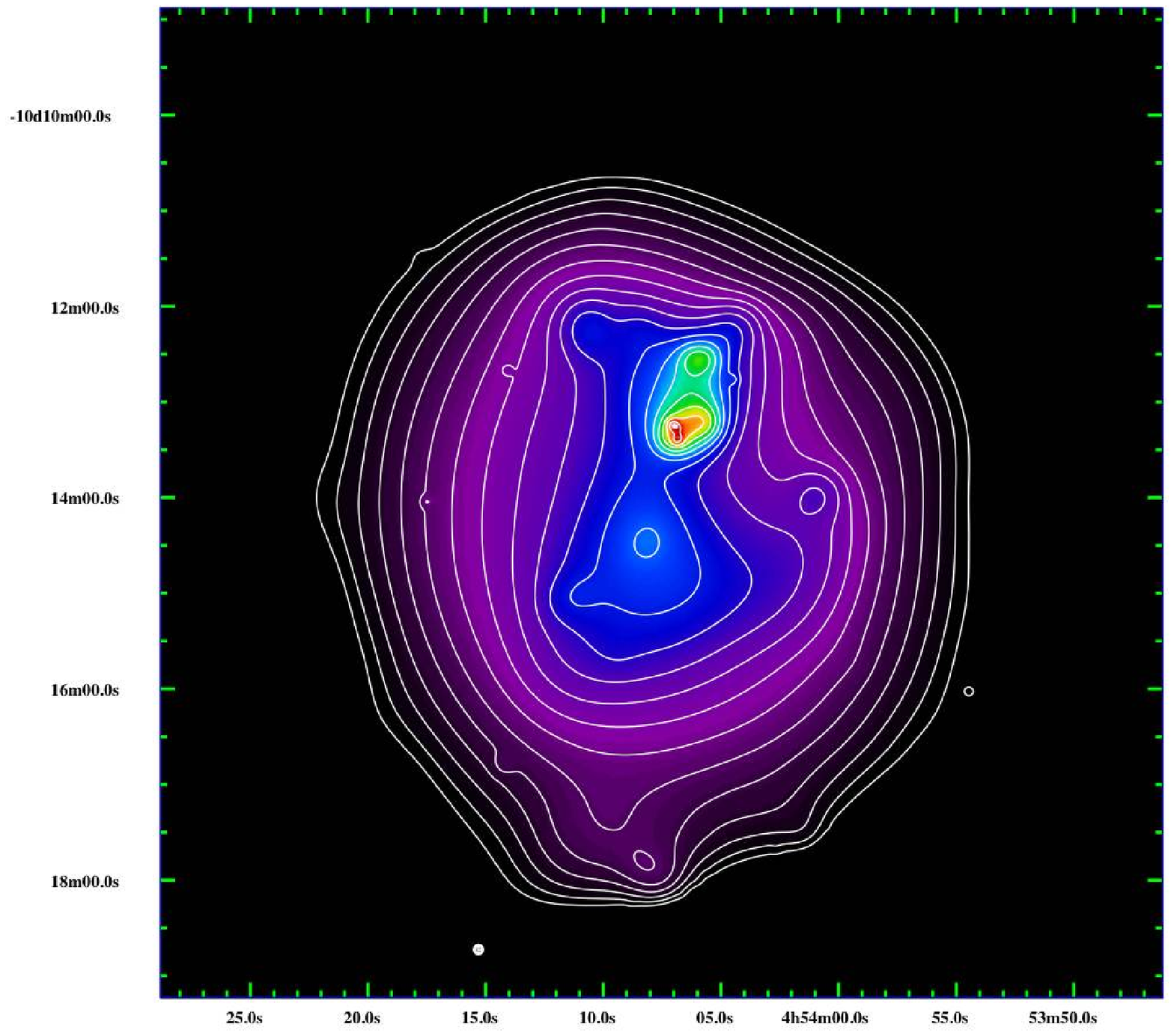}} \hfill \parbox[b]{18cm}{
\caption{0.5-5 keV image of A521 diffuse emission ($7{\times}7~{\rm
arcmin}^2$ central field). The image, exposure corrected and
background subtracted, is the sum of ACIS-I and ACIS-S observations,
and it has been filtered using a wavelet transform with a detection
level of 3.7 $\sigma$. The overlaid isocontours are logarithmically
spaced by a factor of 0.1.}
\label{morf1}} \end{figure*}

\begin{figure*} 
\centering
\resizebox{10cm}{!}{\includegraphics{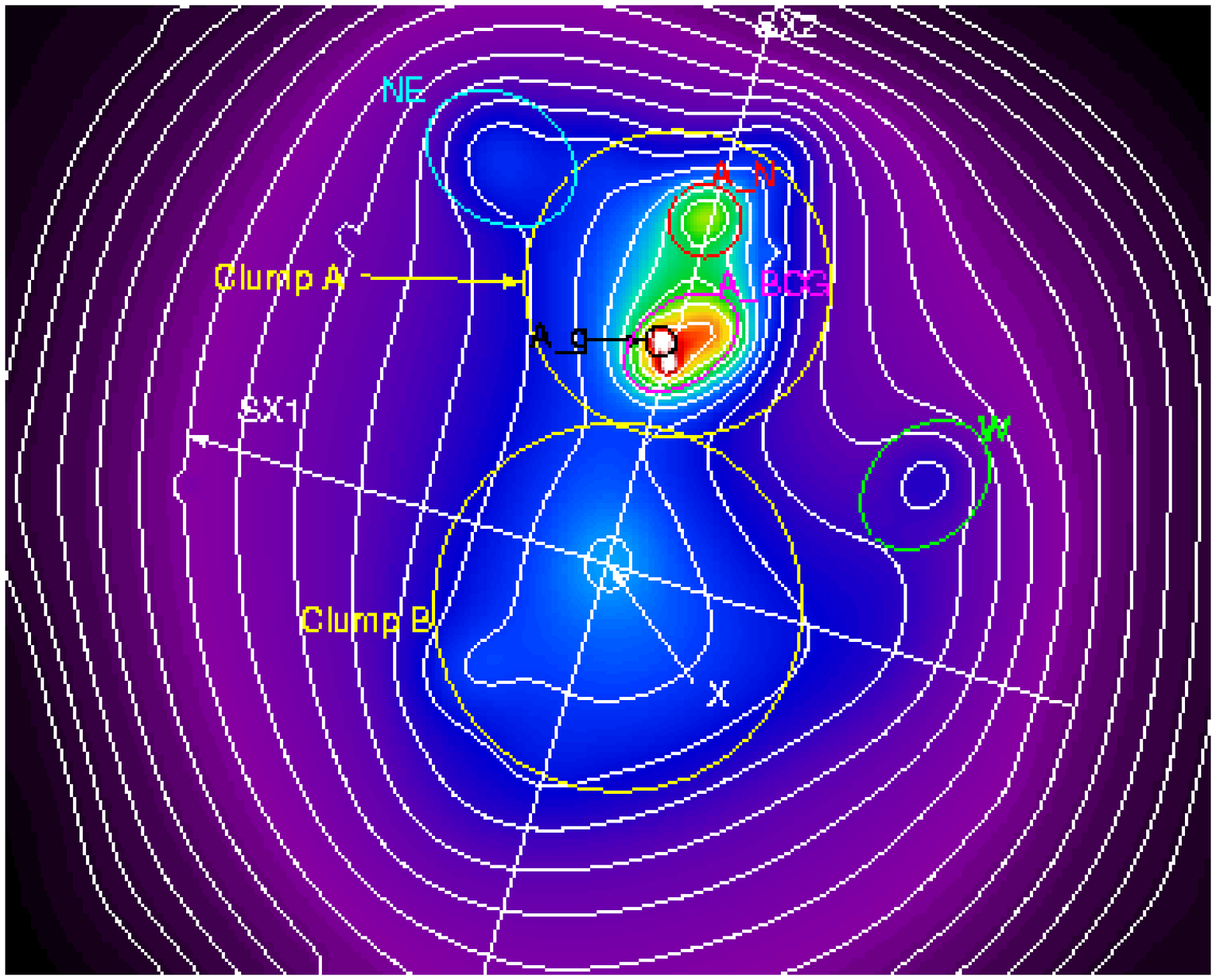}}
\resizebox{10cm}{!}{\includegraphics{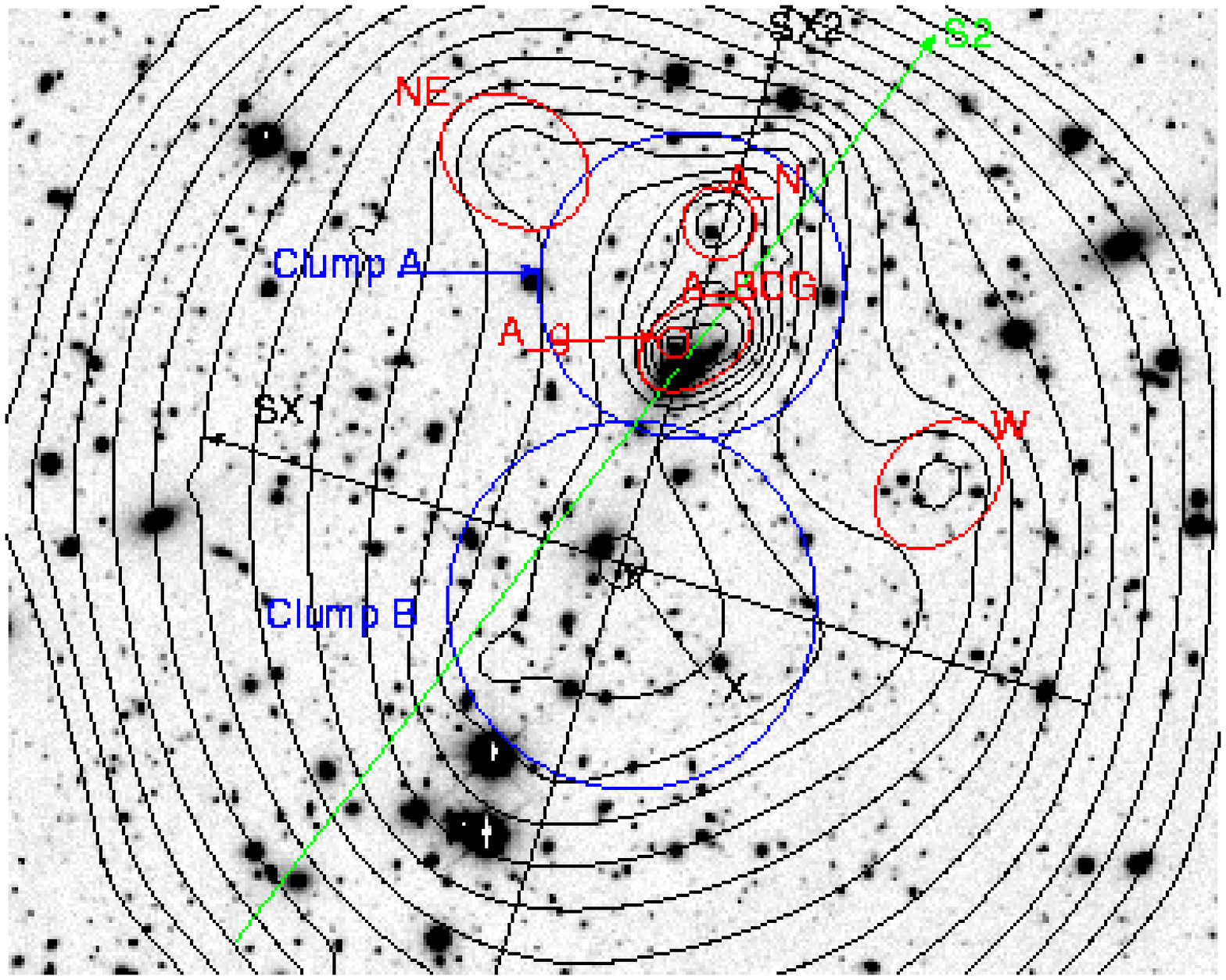}}
\parbox[b]{18cm}{
\caption{{\bf Top:} image of the diffuse X-ray emission of A521 (0.5-5 keV)
- {\bf Bottom:} deep, I-band optical image of the cluster obtained
with CFH12k@CFHT. Contours of the X-ray diffuse emission are overlaid
(logarithmically spaced by a factor of 0.1 as in Fig.~\ref{morf1}).  The
main sub-structures detected in the ICM distribution are indicated by
circles and ellipses: a) the highest density clumps {\it A} (Northern
sub-cluster) and {\it B} (central part of the main cluster) b) the
main sub-structures of clump {\it A}, centred on three X-ray peaks,
{\it A\_BCG} (hosting the BCG), {\it A\_N} (centred on a stellar
object) and {\it A\_g} (corresponding to one of the optical knots that
surround the BCG), and c) the western and north-eastern clumps {\it W}
and {\it NE}. The images cover the $5{\times}5~{\rm arcmin}^2$ central
field of A521. The position of the X-ray peak of the main cluster is
indicated by an {\it X}. The white (top)/black (bottom) line {\it SX2}
joins the X-ray peaks of the main cluster and of the group centred on
the BCG; the corresponding perpendicular direction is indicated with
{\it SX1}. The main axis of the optical distribution (see Ferrari et
al. 2003) is shown by a green line.}
\label{morf2}}
\end{figure*}

The raw image of A521 diffuse emission is presented in
Fig.~\ref{strucKT}. The cluster shows two highest density regions that
we will call clumps {\it A} and {\it B} in the following.  They
correspond respectively to the Northern group and the central part of
the main cluster identified in the ROSAT image (Arnaud et al. 2000).
The new
\cha observations uncover the presence of several features inside and
around each of the two clumps, as it appears more clearly in
Fig.~\ref{morf1}, which represents a smoothed image of the cluster
central field ($7{\times}7~{\rm arcmin}^2$) in the 0.5-5 keV energy
band. In order to get a smoothed image of the diffuse emission of the
cluster, the programme ``mrp\_filter'' of the package ``MR/1
Multiresolution Analysis'' (Stark, Murtagh \& Bijaoui 1998) has been
applied. The programme does a wavelet filtering for images with
Poisson noise. We used a significance level of 1.E-04, corresponding
to a 3.7 $\sigma$ Gaussian detection level. The image has been
thresholded and reconstructed such that both point sources
and the background are excluded, and it has been exposure corrected. In
Fig.~\ref{morf2} the X-ray contours are overlaid on the X-ray (top)
and I-band (bottom) images of the cluster.

The \cha observations reveal a structure of A521 diffuse emission that
is even more complicated and irregular than the morphology obtained
through the ROSAT observations (Arnaud et al. 2000), confirming that
this cluster is out of hydrostatic equilibrium. The general X-ray
structure of A521 is elongated along the axis joining the two main
X-ray peaks ({\it SX2} in Fig.~\ref{morf2}). The green line in the
bottom panel of Fig.~\ref{morf2} shows the direction followed by the
general structure of the cluster at optical wavelengths ({\it S2} in
Ferrari et al. 2003). A deeper optical analysis of the alignment
effects in A521 revealed that the NW/SE direction indicated in green
is the preferred one for the formation of the cluster, since it is the
main elongation axis of a) the brightest cluster galaxies, b) the main
sub-structures detected in the red-sequence iso-density map, and c)
the general cluster structure out to $\sim$ 5${h}^{-1}$ Mpc (Plionis
et al. 2003). A slight misalignment is present between the main axis
of the X-ray and optical emission ({\it SX2} and {\it S2}).

The X-ray peak of the main cluster (labelled {\it X} in Fig.~\ref{morf2})
is close to the second brightest cluster galaxy and its position
corresponds to the barycentre of the optical emission of A521. South
of {\it SX1}, the axis perpendicular to {\it SX2} (see
Fig.~\ref{morf2}), the cluster appears rather relaxed with quite
regular isophotes (Figs. \ref{morf1} and \ref{morf2}).

The Northern part (North of {\it SX1}) is much more complex, as it
shows evidence of several sub-structures and elongations in the ICM
distribution.  The most prominent structure is the clump {\it A} (see
Figs. \ref{strucKT} and \ref{morf2}).  It is brighter than the
equivalent region centred on the main cluster and roughly elliptical
in shape, with a main axis along the {\it SX2} direction. Two internal
sub-structures are detected, a northern clump and a southern one,
labelled respectively {\it A\_N} and {\it A\_BCG} in
Fig.~\ref{morf2}. The brightest peak of clump {\it A}, located inside
the {\it A\_BCG} substructure, is clearly centred on the BCG
position. The substructure {\it A\_BCG} shows an elongation more
toward North-West with respect to the {\it SX2} direction, being in
fact more aligned with the main axis of the optical distribution and
of the BCG ({\it S2} in Ferrari et al. 2003, green line in
Fig.~\ref{morf2}). A compression of the X-ray isophotes is also
observed in the South of the BCG along the {\it S2} direction. The
clump {\it A\_BCG} shows a secondary X-ray peak centred on one of the
blobs of the optical arc-like structure surrounding the BCG ({\it
A\_g} in Fig.~\ref{morf2}). The observations of Ferrari et al. (2003)
have shown that all of these blobs are at the same redshift of the
cluster, and that they could be galaxies falling onto the brightest
cluster object. {\it A\_g} is likely to be a point like emission of
one of these galaxies. The {\it A\_N} structure is centred on a bright
object (a star, based on the spectral analysis by Ferrari et
al. 2003), but its X-ray emission is extended. The northern part of
the whole substructure {\it A} ({\it A\_N} and surrounding regions)
could therefore be a tail of gas of the clump {\it A\_BCG}.

Two other less prominent substructures are present in the North.
First, a North-East clump which appears as an excess of emission east
of clump {\it A} ({\it NE} ellipse in Fig.~\ref{morf2}). Second, we
observe an elongation of the X-ray isocontours on the West side of the
main cluster X-ray peak towards a sub-structure (labelled {\it W} in
Fig.~\ref{morf2}) at $\sim$2.5 arcmin from it in a North-West
direction. Fig.~\ref{morf2} shows that several galaxies are
concentrated in the {\it W} region, i.e. 8 quite bright objects (${\rm
I}_{AB}$=18.5-19) and several faint ones. Spectroscopic observations
reveal that 3 of them are confirmed cluster members, but note that in
this region the spectroscopic observations are complete at no more
than 50-60\% level (Ferrari et al. 2003).

\begin{figure}
\centering
\resizebox{8cm}{!} {\includegraphics{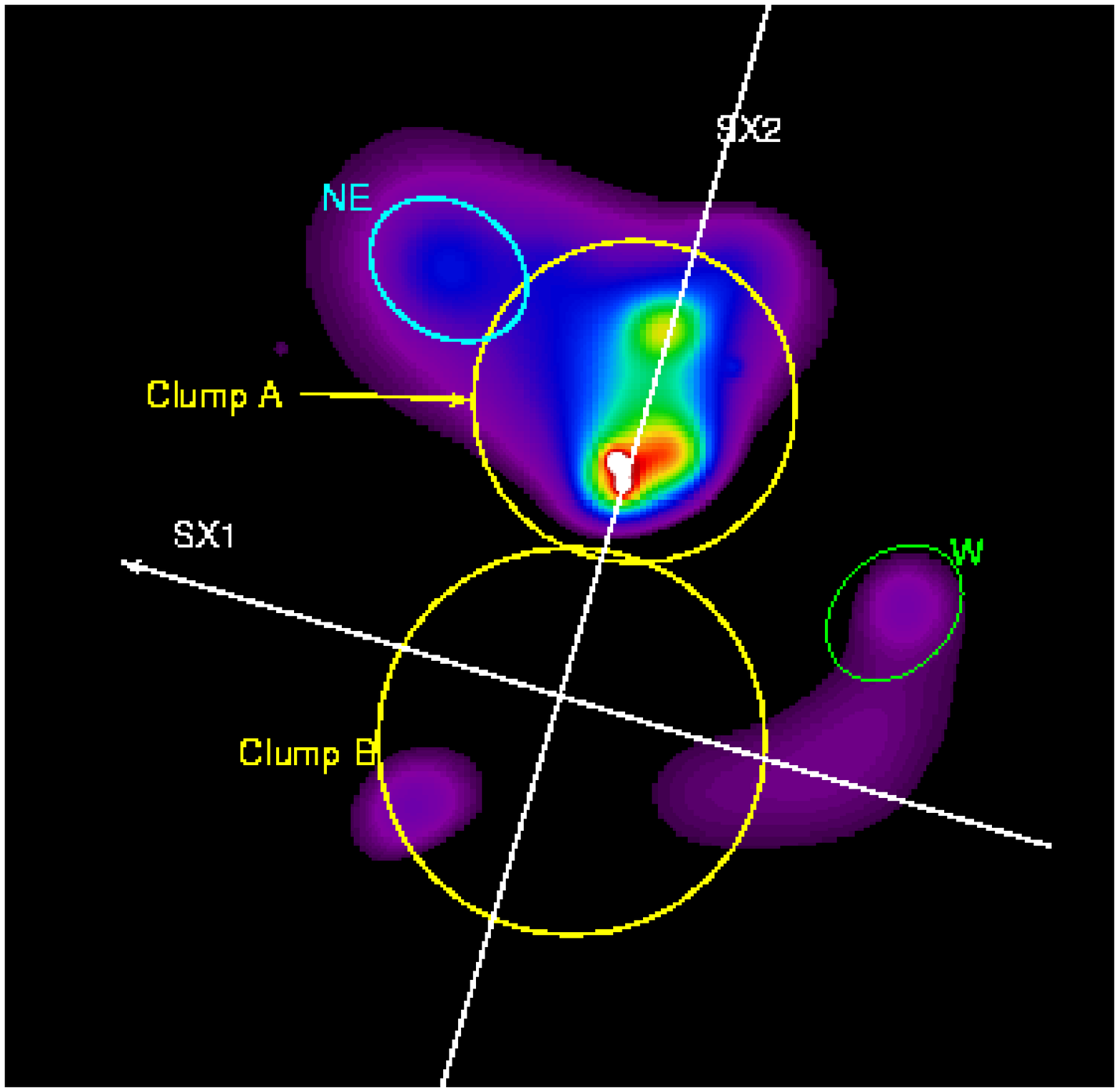}} 
\hfill 
\parbox [b]{8cm}{
\caption{Image of the residuals of A521 X-ray emission: difference
between the image of Fig.~\ref{morf1} and the same image rotated by
180 degrees around the X-ray peak of the main cluster ({\it X} in
Fig.~\ref{morf2}).}
\label{residuals}} 
\end{figure}

The X-ray morphology of A521 is therefore regular South of the {\it
SX1} direction, strongly substructured in the northern part. This is
even clearer in Fig.~\ref{residuals}, which shows the residuals
obtained after subtracting in each region the emission from the
symmetric region with respect to the X-ray peak of the main cluster
(labelled {\it X} in Fig.3).  This way we subtract in the North the
corresponding 'unperturbed' part of the main cluster, as measured in
the South.  Clumps {\it A}, {\it W} and {\it NE} emerge clearly North
of {\it SX1} (with a possible tail of X-ray emission towards
South-East for the substructure {\it W}).

\begin{figure}
\centering
\resizebox{8cm}{!} {\includegraphics{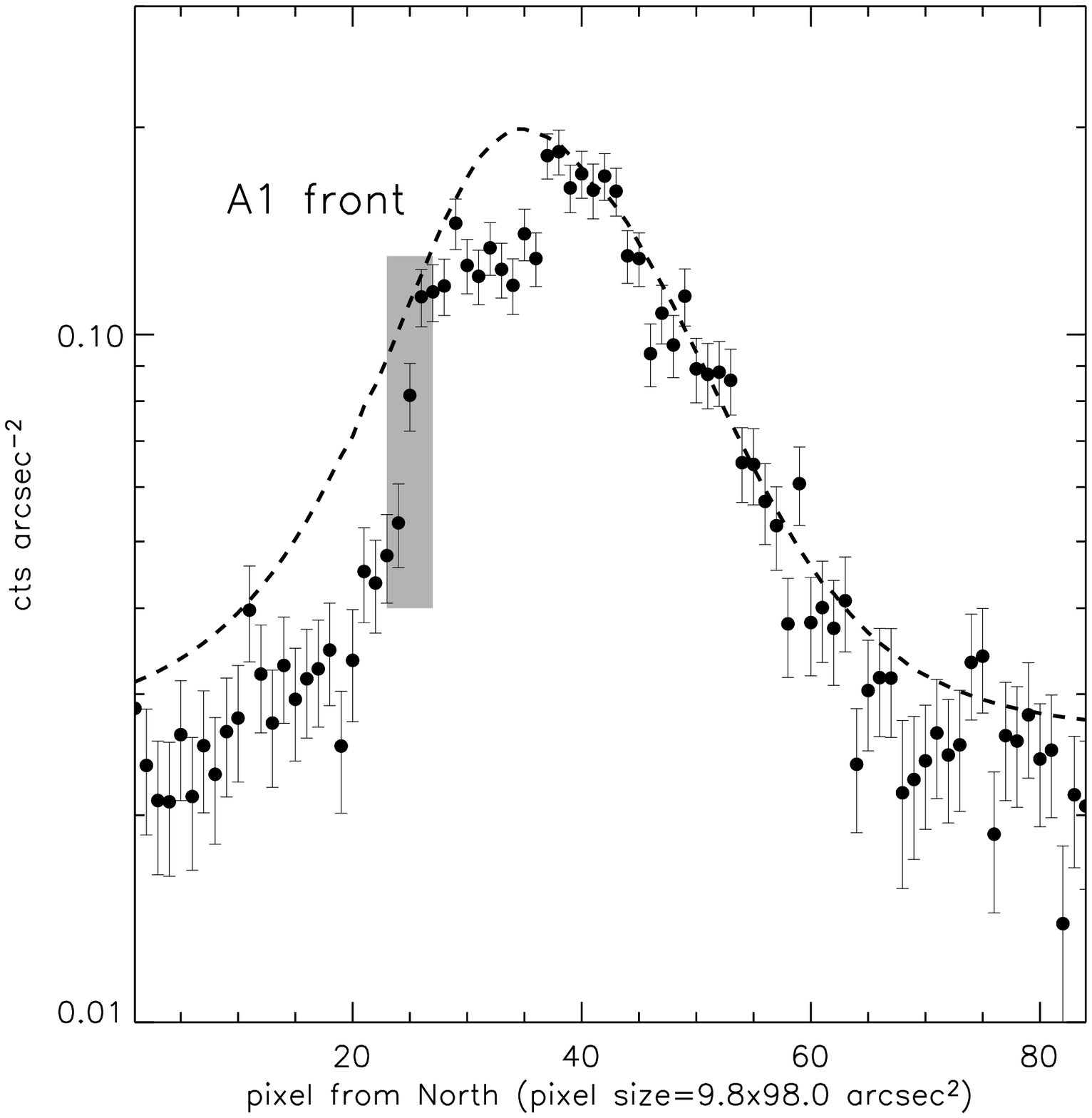}}
\resizebox{8cm}{!} {\includegraphics{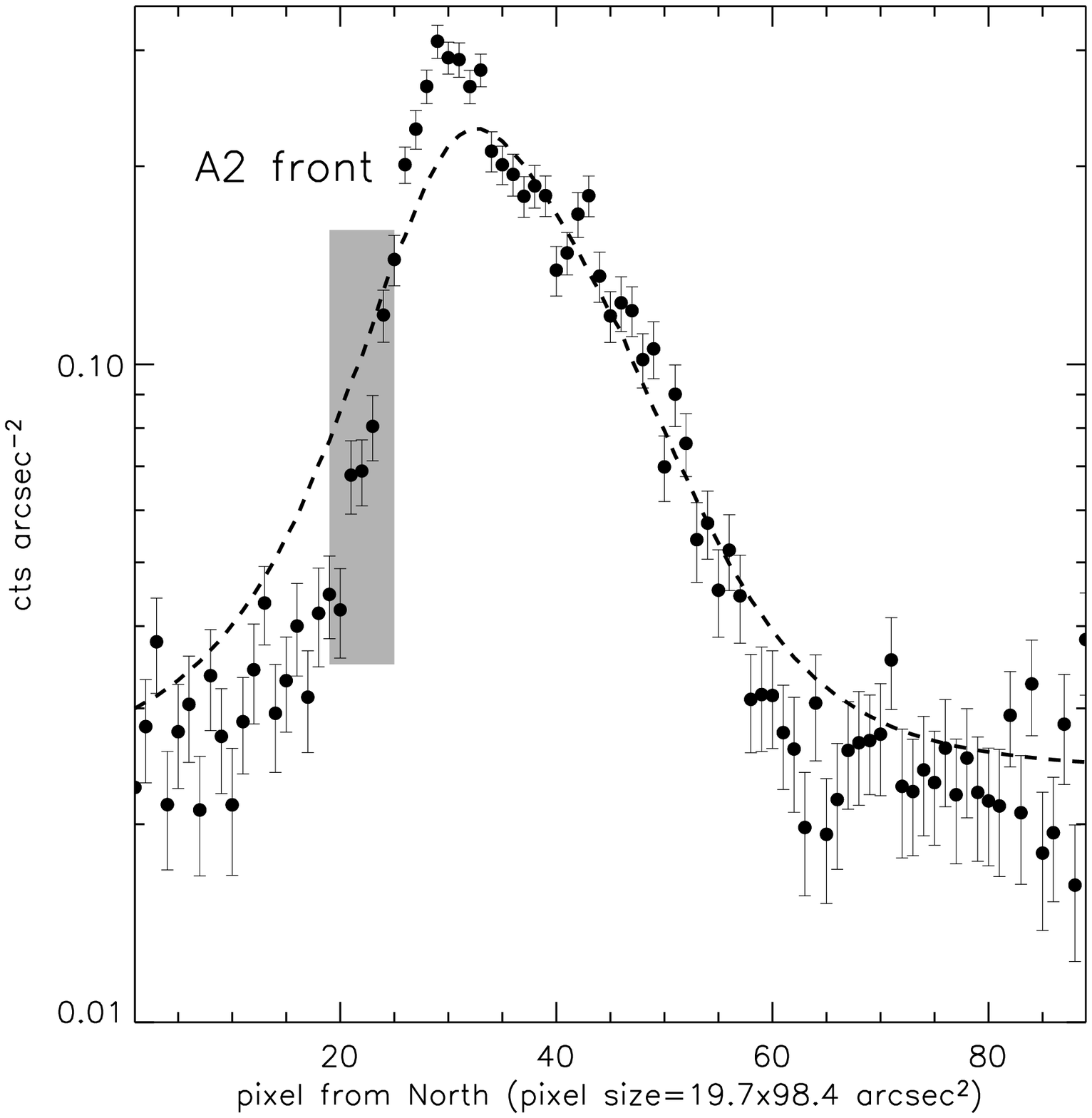}} 
\hfill \parbox[b]{8cm}{
\caption{Counts distribution corresponding to the regions of the two
fronts detected at the North of clump {\it A} (points) and modelled
$\beta$-model from relaxed regions of the cluster (dashed line). The
positions of the two arc-like discontinuities are indicated in grey.}
\label{A1A2}} 
\end{figure}

Finally, \cha observations revealed two sharp edges in the X-ray
surface brightness in the region to the North of the BCG (labelled
{\it A1} and {\it A2} in Fig.~\ref{strucKT}), with changes by a factor
of 2 over scales shorter than $10$ arcsec (see Fig.~\ref{A1A2}).  To
emphasise the departure from a smoothed spherically symmetric
emission, we modelled the surface brightness of the region with two
$\beta-$models, one obtained as best-fit results of the radial profile
extracted from the Northern (P.A.=$[-90^o, 90^o]$) semicircle centred
on the BCG (RA, Dec)=(04:54:06.9, -10:13:20), the second one extracted
from the Southern semicircle centred at (RA, Dec)=(04:54:07.9,
-10:14:27).  Then, we compute the surface brightness profile in two
strips crossing the two edges both in the $(0.5-5)$ keV
exposure-corrected image and in the faked two-dimensional emission
obtained by the sum of the two $\beta-$models. The faked profile is
indicated by dashed lines in Fig.~\ref{A1A2}.  These edges are located
orthogonal to the main X-ray axis {\it SX2} and trace the Northern
boundary of the clump~{\it A} (see Fig.~\ref{strucKT}).

In summary, we distinguish the following main structures in the X-ray
observation of A521:

\begin{itemize}
\item {\it clumps A and B}: the two main clumps in the X-ray emission of
A521. Clump {\it B} is the central part of the main cluster, which
seems to be nearly unperturbed in the southern region. Clump {\it A}
is a northern sub-cluster centred on the BCG;

\item {\it SX1, SX2, S2}: {\it SX2} is the axis connecting the X-ray
peaks of the main cluster and of the northern sub-cluster  (with a
position angle of 163$^\circ$\footnote{The position angle being
defined from North to East.}), {\it SX1} its perpendicular direction,
{\it S2} the optical main elongation axis;

\item {\it A\_BCG}, {\it A\_N} and  {\it A\_g}: X-ray structures  inside the
Northern sub-cluster {\it A}.  {\it A\_BCG} is a centred on the BCG
galaxy and {\it A\_g} is associated with one of the optical knots
surrounding the BCG;

\item {\it W}: substructure present at 2.5 arcmin North-West of the X- ray
peak of the main cluster;

\item {\it NE}: substructure located North-East of clump {\it A};

\item {\it A1} and {\it A2}: two arc-like edges detected as in the X- ray 
brightness map at the North of the clump {\it A}.

\end{itemize}

The spherical symmetry and hydrostatic equilibrium assumptions are
clearly not valid in the case of A521 due to its very disturbed
morphology. No gas density and temperature profiles or total mass
profiles will therefore be presented in the following.

\subsection{Temperature analysis}

\subsubsection{Temperature map from hardness ratio}

\begin{figure*}
\centering
\resizebox{18cm}{!}{\includegraphics{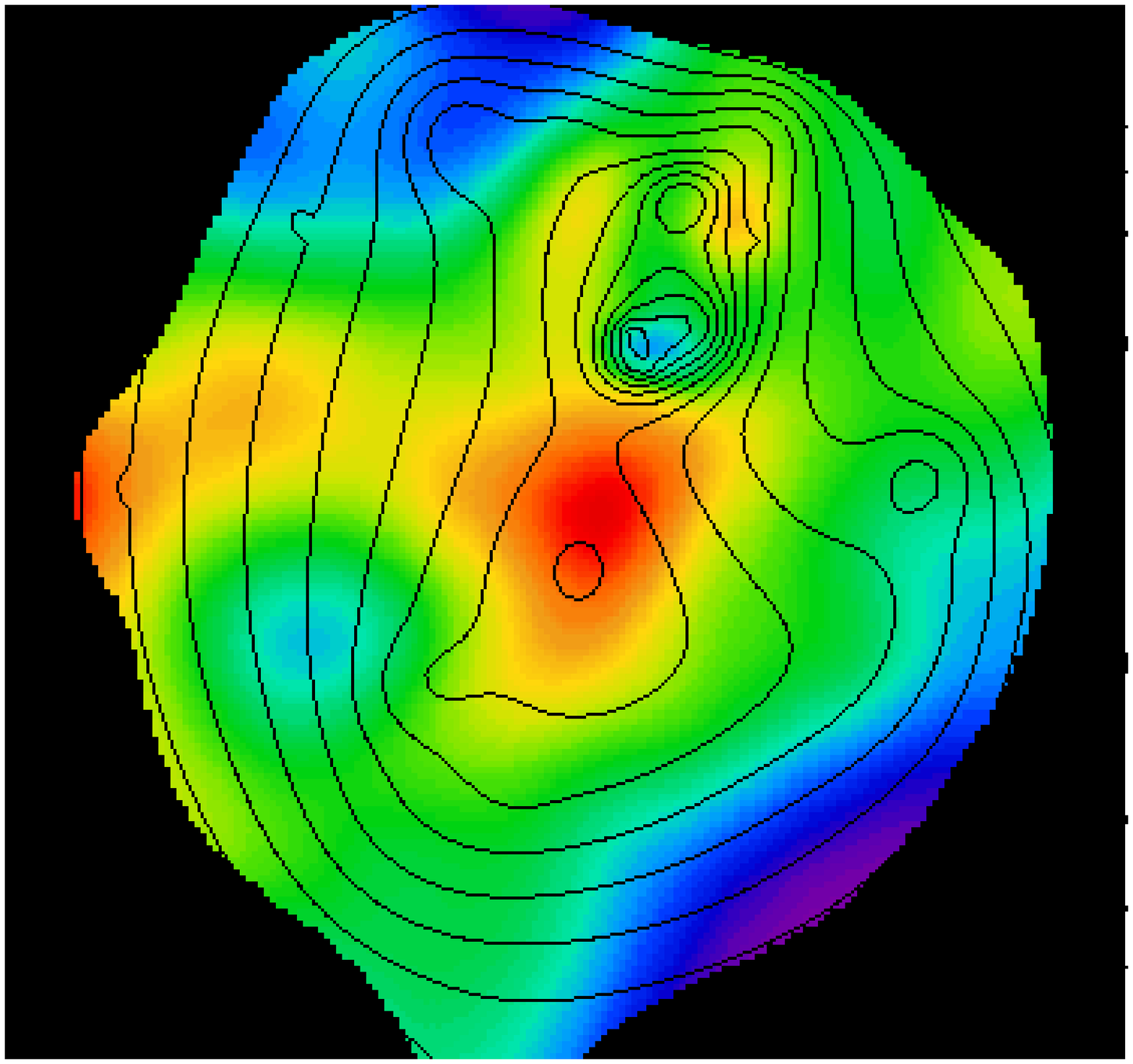}{\includegraphics{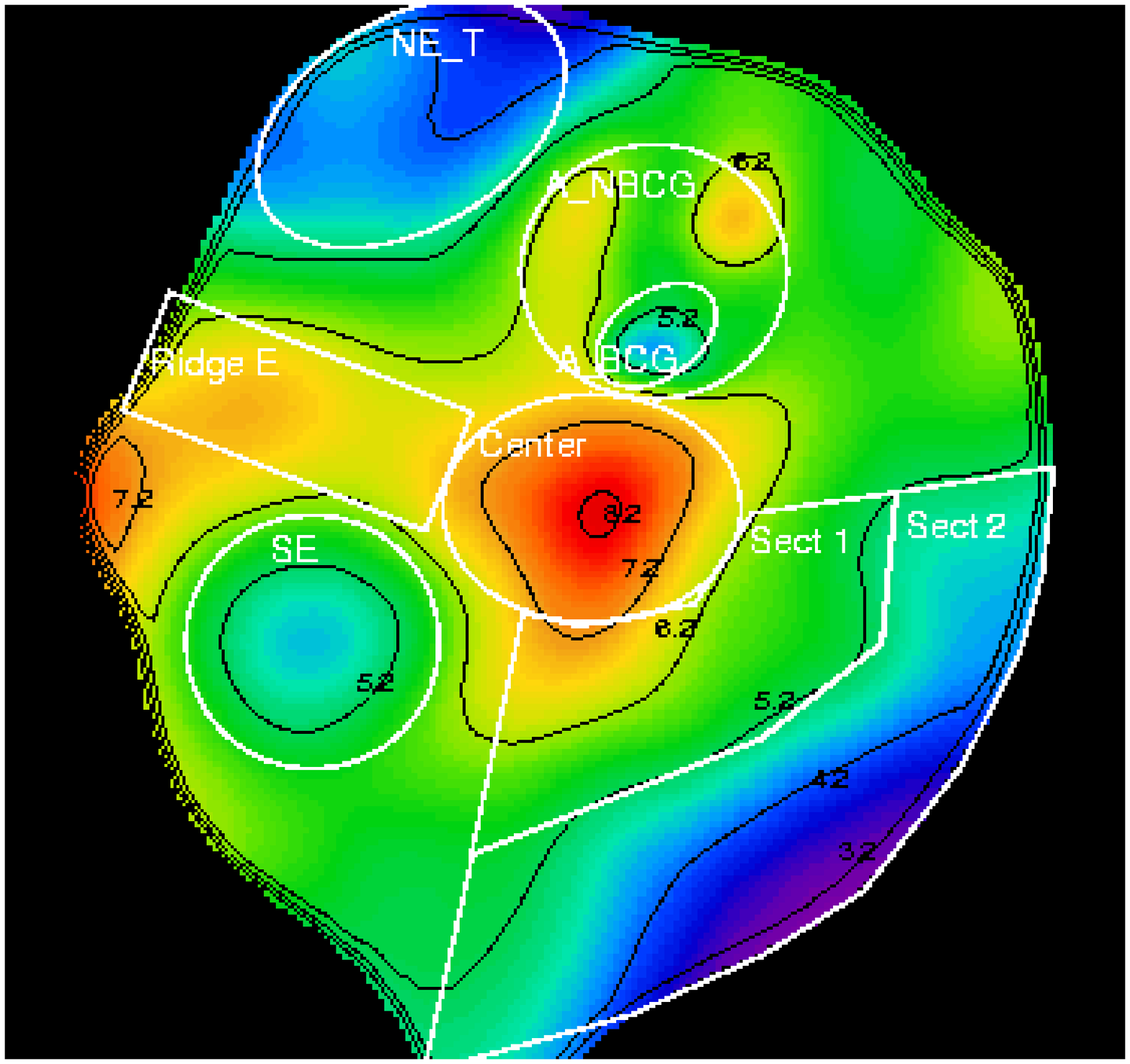}}}
\hfill  
\parbox[b]{18cm}{
 \caption{Temperature map obtained through the hardness ratio
 technique (see text for details). The colours range from purple
 ($\sim$ 3.5 keV) to red ($\geq$ 8 keV). {\bf Left:} the isocontours of
 the X-ray diffuse emission of A521 are superposed (logarithmically
 spaced by a factor of 0.1 as in Fig.~\ref{morf1}). {\bf Right:} the
 temperature isocontours and the corresponding k$T$ values are
 overlaid in black. The regions used for the spectroscopic analysis
 are superposed in white. }
\label{hr}}  
\end{figure*}

Fig.~\ref{hr} shows a temperature map of A521 obtained using the
hardness ratio technique. Images of the region covered by the ACIS-S3
field of view have been extracted in the energy bands 0.5-2 keV and
2-5 keV from the ACIS-S and ACIS-I event files.  Point sources have
been detected and removed using the CIAO tools {\sc wavdetect} and
{\sc dmfilth}.  Each image has been background subtracted using blank
field data and corrected for vignetting effect and exposure
variations. The resulting ACIS-S and ACIS-I count rate images in each
energy band have then been added and adaptively smoothed using the
CIAO tool {\sc csmooth}. The smoothing scales were defined from the
raw ACIS-(S+I) image in the $ 2-5$ keV energy band with a minimum
significance of $4~\sigma$ and a maximum significance of $5~\sigma$ .

The 2-5 keV smoothed image has been divided by the corresponding 0.5-2
keV image to obtain a hardness ratio map, which has been converted to
a temperature map.  The theoretical conversion factors have been
computed using an absorbed thermal model ({\tt tbabs (mekal)} in XSPEC
11.3.1) with a column density fixed to the Galactic value of
$5.79{\times}10^{20}$ cm$^{-2}$ , a redshift of 0.247 and an abundance
fixed to 0.4, convolved with the instrument responses. Since the
images have been corrected for vignetting effects, we used the on-axis
Auxiliary Response and Redistribution Matrix files, obtained
accordingly to the period and configuration of the observation.  An
inverse edge to account for the underestimate of the effective area
around 2 keV is also applied (see Vikhlinin et al. 2005).

A521 is clearly characterised by a highly sub-structured temperature
map, which presents:

\begin{itemize}

\item [-]  a cold  region ($T\leq$5 keV) in the North-East (labelled 
{\it NE\_T}) around the {\it NE} substructure;

\item [-] a cold substructure  corresponding to the
region of the BCG ({\it A\_BCG}). It is surrounded by an annulus of
warmer gas ({\it A\_NBCG}), that shows higher temperatures in its
northern and eastern parts;

\item [-] a central hot region ($\sim$6 to 8 keV) that, starting from  East,
runs roughly parallel to the SX1 direction ({\it Ridge E}) and reaches a
maximum in a very hot central peak ({\it Central});

\item [-]  a gradual decrease of temperature in the South-West  sector  
({\it Sect 1} and {\it Sect 2}). The temperature gradient is less
pronounced in the South-East sector, with a possible cold
substructure South-East of the cluster barycentre ({\it SE}).

\end{itemize}

\subsubsection{Spectral analysis}

\begin{table*}
\begin{center}   \begin{tabular}{ccccccc}  \hline
\hline
Region & $T$ & $Z$ & $f_X$(0.5-2 keV) & $f_X$(bol) & $L_X$(0.5-2 keV)
& $L_X$(bol) \\ 
       & (keV) & ($Z_{\odot}$) & ($10^{-12}$ erg/s/${\rm
cm}^2$) & ($10^{-12}$ erg/s/${\rm cm}^2$) & ($10^{44}$ erg/s) &
($10^{44}$ erg/s) \\
\hline
Whole cluster & $5.85{\pm}0.23$ & $0.55 {\pm}0.08$ & $2.03{\pm}0.03$ & $6.47{\pm}0.13$ & $3.42{\pm}0.05$ & $12.94{\pm}0.21$ \\
Clump A & $5.40{\pm}0.50$ & $0.50{\pm}0.22$ & $0.35{\pm}0.02$ & $1.06{\pm}0.07$ & $0.60{\pm}0.04$ & $2.16{\pm}0.13$ \\
Clump B & $5.79 {\pm}0.56$ & $0.40 {\pm}0.21$ & $0.38{\pm}0.02$ & $1.18{\pm}0.08$ & $0.64{\pm}0.04$ & $2.38{\pm}0.14$ \\
\hline
\end{tabular}
\caption{The results of spectral fitting in the whole cluster (circular region 
centred at (RA, Dec)=(04:54:07.9, -10:14:42.6) and with a radius of 4 arcmin)
and in the two main clumps A and B (circular regions centred
respectively at (RA, Dec)=(04:54:06.6, -10:12:56.3) and (RA,
Dec)=(04:54:07.9, -10:14:42.6), and with radius of 0.8 and 1.0
arcmin). $1\sigma$ errors are given. $N_{\rm H}$ has been fixed to the
galactic value (5.79${\times}10^{20} {\rm cm}^{-2}$).}
\label{tab:spectral}
\end{center}
\end{table*}

Spectra in different regions of interest were extracted, together with
the corresponding Auxiliary Response and Redistribution Matrix files.
As above, an absorbed thermal model is used to fit the accumulated
spectra in bins with a minimum number of 20 counts, with a column
density fixed to the Galactic value. An inverse edge to account for
the underestimate of the effective area around 2 keV is also applied
(see Vikhlinin et al. 2005).

The temperatures, abundances, fluxes and rest-frame luminosities of
the whole cluster (big circle in Fig.~\ref{strucKT}\footnote{The CCD
gaps have been masked in doing the spectral analysis of the whole
cluster.}) and of the two main clumps {\it A} and {\it B} (smaller
circles in Fig.~\ref{strucKT}) are listed in
Table~\ref{tab:spectral}. The quoted values have been derived using
ACIS-I data, for which a local background can be estimated. They do
not show significant variation when blank field background are used,
validating the use of blank field data for the temperature map.

The spectroscopic temperatures ($T_{\rm spec}$) of the different
regions identified in the temperature map (see previous section) have
also been estimated. The ACIS-I and ACIS-S spectra were fitted
simultaneously
\footnote{We checked that fully consistent results are obtained using  the 
ACIS-I exposure only}.  These temperatures are compared in
Table~\ref{tab:specTOT} with the temperatures obtained using the
hardness ratio technique. The latter ($T_{\rm HR}$) have been
estimated using both the mean value from the temperature map and the
ratio of the count rates in the smoothed images used to derive the
temperature map, these two methods giving results which always differ
by less than $0.1$ keV.

The temperatures derived from the spectral analysis are in good
agreement, within 1$ \sigma$ error, with the value derived from the
hardness-ratio technique (see Table~\ref{tab:specTOT}), with no
systematic differences.  The largest discrepancy ($\sim 1.1\sigma$) is
observed for the region around the BCG ({\it A\_BCG}) which is even
colder in the spectroscopic analysis. The spectral analysis clearly
confirms: a) the low temperature of the clump centred on the BCG,
which is surrounded by a hot region, b) the presence of a hot central
bar elongated in a East/West direction with a maximum in the central
region, and c) the cold temperature of the North-East part of the
cluster.

\begin{table}
\begin{center}   \begin{tabular}{ccc}  \hline
\hline
Region & $T_{spec}$ & $T_{\rm HR}$ \\ & (keV) & (keV)\\
\hline
{\it NE\_T} & $4.10{\pm}0.66$ & 4.5\\
{\it A\_BCG} & $4.25{\pm}0.80$ & 5.1 \\
{\it A\_NBCG} &  $6.51{\pm}0.89$ & 6.0  \\
{\it Ridge E} & $5.26{\pm}1.75$ & 6.6 \\
{\it Center} &  $7.44{\pm}1.19$ & 7.4 \\
{\it Sect 1} &  $5.77{\pm}0.68$ & 5.9 \\
{\it Sect 2} &  $4.77{\pm}0.66$ & 4.6 \\
{\it SE} &  $4.70{\pm}0.88$ & 5.3 \\
\hline
\end{tabular}
\caption{Spectroscopically derived temperatures, $T_{spec}$, of  the 
different structures detected in the temperature map of A521 (see
Fig.~\ref{hr}). $T_{spec}$ has been obtained by fitting simultaneously
the ACIS-I and ACIS-S spectra.  All the quoted results are obtained
with $N_{\rm H}$ fixed to the galactic value, solar abundance Z=0.4 as
in Anders \& Grevesse (1989), and provide a reduced $\chi2 <1$.  The
second column gives the temperature derived using the hardness ratio
technique.}
\label{tab:specTOT}
\end{center}
\end{table}

\section{Discussion}\label{disc}

In agreement with previous results by Arnaud et al. (2000),
Maurogordato et al. (2000) and Ferrari et al. (2003), the analysis of
\cha observations confirms that A521 is far from dynamical equilibrium
and that it is a particularly complex system, made up by several
subclusters in different phases of the merging process.  A first
evidence that A521 is in a disturbed dynamical state is the slight
misalignment between the main axis of X-ray and optical emissions
({\it SX2} and {\it S2} in Fig.~\ref{morf2}). This could be due to
on-going merging event(s), since we know that in clusters the
collisional component (i.e. ICM) and the non-collisional one
(i.e. galaxies and DM) have significantly different dynamical time
scales (R\"ottiger et al. 1993).

\subsection{BCG group and surrounding regions}

While the Southern part of A521 has a quite regular morphology and 
shows a peak of X-ray emission very close to the second BCG, the
Northern region has a much more structured X-ray morphology, rich in
sub-clusters.  The main feature that appears in its ICM density and
temperature maps is a compact cold component (clump {\it A\_BCG},
Fig.~\ref{morf2}), centred on the BCG, which is the highest density
region of the principal northern substructure (clump {\it A}).  The
very high angular resolution of
\cha has helped to solve one of the open questions of the previous
X-ray analysis of A521 (Arnaud et al. 2000). In the ROSAT observations
a compact group in the North of the cluster was also detected, hosting
the BCG. Curiously, it was not centred on its brightest galaxy, but on
a northern position. \cha observations show two main X-ray peaks in
the subcluster {\it A} corresponding to the ROSAT group, one centred
on a stellar object ({\it A\_N}) and one associated to the BCG ({\it
A\_BCG}), which were not resolved by the previous X-ray
observations. The {\it A\_BCG} substructure is perfectly centred on
the BCG and it corresponds spatially to the gravitationally bound
system of galaxies detected by Ferrari et al. (2003). The gas of this
clump, significantly hotter than the diffuse ISM of early type
galaxies ($\sim$0.5-1.5 keV, Forman et al. 1985), is associated to the
whole system of galaxies surrounding the BCG. Its temperature, colder
than in the rest of the cluster, is in agreement with the lower radial
velocity dispersion of this region ($256 ^{+82}_{-133} {\rm km}{\rm
s}^{-1}$ in a circle of 240 ${h_{75}}^{-1}$ kpc around the brightest
galaxy, Ferrari et al. 2003). What we need to investigate is the
origin of the high gas density in {\it A\_BCG}, $\sim$1.5 times higher
than in the centre of the main cluster\footnote{The density ratio has
been measured as the square root of the surface brightness ratio. The
0.5-5 keV surface brightness maps (ACIS-I and ACIS-S) have been used,
considering the {\it A\_BCG} region and an equivalent region around
the X-ray centre of the main cluster.}.

Similar to 28\% of the first-rank galaxies in rich clusters (Hoessel
1980), the brightest galaxy of A521 is clearly a case of BCG with
multiple nuclei (see Fig.~8 of Maurogordato et al. 2000). Since all of
them are at the same redshift (Ferrari et al. 2003), they are very
likely remnants of galaxies ``cannibalised'' by the most massive
object, around which, however, the halo typical of cD galaxies has not
been detected (Maurogordato et al. 2000). The formation of the halo in
cD galaxies seems to result from the tidal disruption of a large
fraction of dwarf galaxies during the early stages of cluster
evolution (Merritt 1984; L\'opez-Cruz et al. 1997). Subsequently,
violent relaxation redistribute the stars from the disrupted objects
throughout the cluster's potential, giving rise to the cD's halo,
while the gas originally confined in the cannibalised galaxies can
contribute significantly to the ICM mass (L\'opez-Cruz et
al. 1997). In A521 we are probably observing the initial phase of the
formation of a cD at the centre of a low-mass subcluster, in which
galaxy merging is efficient due to the low velocity dispersion of the
system, while the extended halo has not yet had time to form.  The
very high ICM density of the {\it A\_BCG} group could therefore be due
to stripped material related to the cannibalism of the BCG. Galaxy
cannibalism occurs very early in a cluster lifetime. If it would
happen after cluster virialization, we should expect randomly oriented
cD galaxies, while it has been shown that the shape of cD's aligns
with its nearest neighbour, the cluster shape and the filaments of
large scale structure (West 1994). The scenario that the BCG of A521
is becoming a cD galaxy in a dynamically young group is therefore
supported by the observed alignment of the BCG main axis with the
cluster major axis and with the nearest cluster neighbour (Plionis et
al. 2003). It is also in agreement with the idea that cD's form via
the merging of galaxies in the centre of poor groups, which then fall
into richer clusters (Merritt 1984; Zabludoff \& Mulchaey 1998 and
references therein). Notice that a similar case of a cold and very
dense clump of gas, detected around a BCG with multiple nuclei and
aligned along the merging axis of the cluster, has been observed with
\cha in A3266 (Henriksen \& Tittley 2002). In that case, the BCG is
probably cannibalising galaxies from a merging subcluster.

In agreement with the previous optical analysis, we therefore conclude
that the main northern clump {\it A} is a group of galaxies in
interaction with the main cluster. The subcluster shows a higher
density component centred on the BCG, a northern tail of gas, and
compression of the X-ray isophotes South of the BCG. All these results
are in agreement with the merging scenario suggested by our optical
analysis of the cluster (i.e. the group {\it A} is infalling toward
the main cluster along a North/South direction). By comparing the ICM
temperature and density maps of A521 with the numerical simulations of
Ricker \& Sarazin (2001), we suggest that the group {\it A} and the
main cluster are in a pre-merger phase ($\sim$--0.5~Gyr from the
closest cores encounter), with a quite low impact parameter ($\lesssim
1-2 r_s$) and a merger axis nearly perpendicular to the line of sight.

\begin{figure}  
\centering
\resizebox{8cm}{!}{\includegraphics{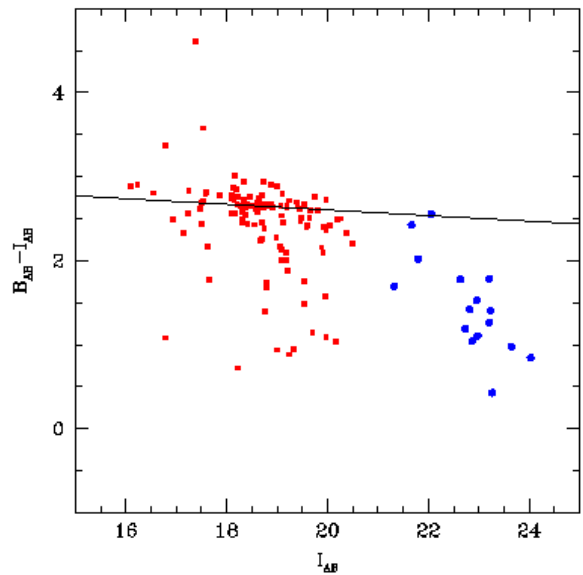}}
\parbox[b]{8cm}{
\caption{Colour-magnitude diagram for the confirmed cluster members (squares) 
and for the galaxies in the {\it NE} substructure detected on the X-ray
map of A521 (circles). The solid line shows the red-sequence best fit
(Ferrari et al. 2003).}
\label{CMDNE}}
\end{figure}

A higher ICM temperature has been observed in the region surrounding
the clump {\it A\_N}, and in particular in its North and North-East
sides (see Fig.~\ref{hr}). 
%One hypothesis is that this is due to the
%ongoing merging event between the northern clump {\it A} and the main
%cluster, which is heating the gas of the lowest mass component ({\it
%A}) with the exception of (or at least less efficiently in) its
%densest and coldest central part ({\it A\_BCG}). 
This higher temperature could be related to the presence of another
substructure observed in the ICM density map of A521, i.e. the
North-East clump {\it NE}. This substructure could be a dynamically
separated group of galaxies, since it shows a low gas temperature and
it hosts some faint galaxies (Fig. \ref{morf2}). The clump might be:
a) another infalling sub-structure at the redshift of A521, or b) a
background gravitationally structure, by chance seen in projection
nearby the central field of A521. In the first case, the low ICM
temperature of this region suggests that the possible interaction
would be in its very initial phase. Several galaxies are present in
the {\it NE} clump, for which we have no redshift information due to
their very faint magnitudes (${\rm B}_{AB}>$23.5, ${\rm I}_{AB}{\geq}$21.5). 
Their positions on the ${\rm (B-I)}_{AB} vs. {\rm
I}_{AB}$ diagram (circles in Fig.~\ref{CMDNE}) do not exclude that the
{\it NE} clump could actually correspond to a group of galaxies at the
distance of A521, since some objects (i.e. the brightest ones, ${\rm
I}_{AB}{\simeq}21.5$) lie on the cluster red-sequence. They could
therefore be faint ellipticals at the cluster redshift, surrounded by
bluer late-type galaxies. It is however not clear why all the galaxies
of this group are fainter than the confirmed members of A521 if they
are at the same redshift. The low temperature of the gas in the {\it
NE} region could also be in agreement with the second hypothesis,
i.e. the {\it NE} clump could be a background group of galaxies. This
seems however less probable, since the faint galaxies located in the
clump lie on or are even bluer than the cluster red-sequence
(Fig.~\ref{CMDNE}), while they should be redder if they were massive
ellipticals at higher redshift. They could be a grouping of late-type
background galaxies, but in such a case the system would not be
massive enough to have such a strong X-ray emission, and therefore it
would not be associated to the {\it NE} clump detected by {\it
Chandra}. Due to the faint magnitudes of these galaxies and the
consequent bigger errors in their colour and magnitude determination,
it is however not possible to exclude that they are elliptical
galaxies at a higher but not so different redshift than A521
(${\Delta}z\sim$0.1). With the optical observations of A521 available
at present it is therefore impossible to give a definitive conclusion
on the nature of the clump {\it NE}.

\subsection{Central hot region}

\begin{figure}  
\centering
\resizebox{8cm}{!}{\includegraphics{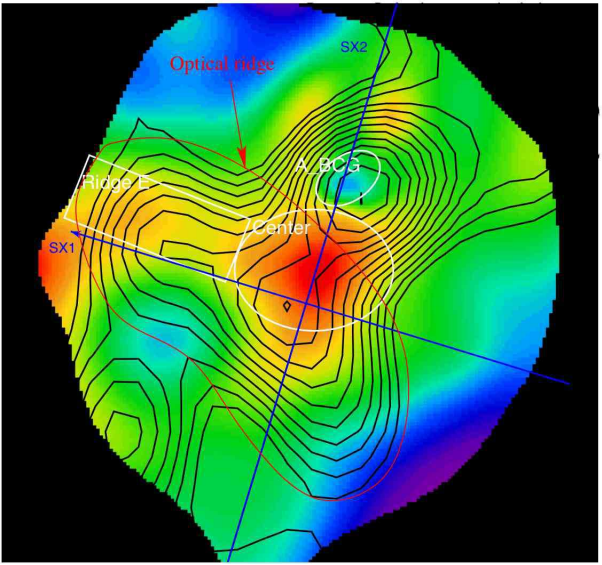}}
\hfill
\parbox[b]{8cm}{
\caption{Iso-density contours of the galaxies with $B<27$ and $I<20$ 
superimposed in black on the temperature map of A521. The over-dense
ridge of galaxies is indicated in red.}
\label{T_IsoD}}
\end{figure}

\begin{figure}  
\centering
\resizebox{8cm}{!}{\includegraphics{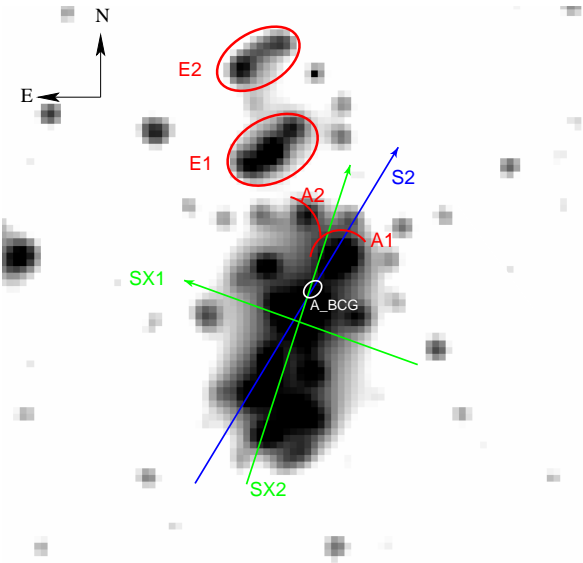}}
\parbox[b]{8cm}{
\caption{Projected galaxy density map of the red sequence galaxies in A521 
central field ($15{\times}20~{\rm arcmin}^2$). The map has been built
on the basis of a multi-scale approach (see Ferrari et al. 2005 for
more details). The positions of the {\it A\_BCG}, {\it A1} and {\it
A2} features are shown. The {\it SX1}, {\it SX2} and {\it S2}
directions are also indicated. Recent observations have confirmed the
presence of several galaxies at the redshift of A521 in the two clumps
{\it E1} and {\it E2}.}
\label{LSS}}
\end{figure}  

Between the main cluster and clump {\it A} we detect a very high
temperature bar roughly parallel to {\it SX1} and extending from East
to the centre of the cluster, where the ICM temperature reaches its
maximum value (Fig.~\ref{hr}). The hot bar corresponds to the eastern
part of the over-dense ridge of galaxies detected by Arnaud et
al. (2000) and Ferrari et al. (2003) on the galaxy iso-density maps
(Fig.~\ref{T_IsoD}). In optical, the ridge bends towards South-West on
the western side of the cluster centre, while the hot bar does not
extend in this western region.

The dynamical properties of this part of the cluster were interpreted
as the result of a recent merger in the ridge region with a
significant component along the line of sight (Ferrari et
al. 2003). The high temperature of the eastern side of the ridge, due
to gas compression and heating, could therefore be associated either
a) to the on-going infall of the subcluster {\it A} toward the centre
of the main cluster, or b) to the merging event nearly along the line
of sight detected in optical. The first hypothesis is in agreement
with the results of numerical simulations, which, in the case of low
impact parameters, show a high temperature bar nearly perpendicular to
the collision axis during the pre-merger phases (Schindler \& M\"uller
1993; Takizawa 1999; Ricker \& Sarazin 2001). Considering hypothesis
b), we can exclude that we are observing the central phases of a
subclusters' collision: due to the merging geometry of the
substructures in the ridge reconstructed through the optical
observations, we should not detect such a high temperature bar (collision
axis nearly along the line of sight prevents to observe strong
signatures of interactions in the X-ray temperature and density maps,
e.g. Schindler \& M\"uller 1993). The high temperature could be
explained by the b) hypothesis in the case of a post-merger, since we
could then be observing in projection the shock fronts moving
outwards. Of course, the hot ridge in the centre of the cluster could
be due to a combination of the two merging events, i.e. the pre-merger
phase of the clump {\it A} and the post-collision nearly along the
line of sight along the ridge.

\subsection{Other features in the density and temperature maps}\label{other}

Other features have been detected in the northern part of the cluster
(i.e. North of the {\it SX1} direction). First of all, the western
clump {\it W}. Optical observations reveal that it hosts some bright
cluster members (bottom panel of Fig.~\ref{morf2}); it could therefore
be a dynamically bound group of galaxies at some stage of interaction
with the main cluster. Since the ICM temperature is higher in the
North than in the South of the group, and the map of the residuals
(Fig.~\ref{residuals}) shows a tail of gas elongated in the SE
direction, we could be witnessing a merging event between the main
cluster and the clump {\it W}, the latter coming from somewhere in the
S-SE direction. An off-axis collision could have prevented the total
assimilation of the less massive group {\it W} in the main component
of A521.

Two edges {\it A1} and {\it A2} have been detected in the X-ray
surface brightness of A521. They are located orthogonally to, and at
opposite directions with respect to, the merging axis {\it SX2} and
trace the Northern boundary of the clump~{\it A}. In the scenario in
which the {\it A\_BCG} group is falling from North onto the main
cluster, the edges might be interpreted as residuals of the sloshing
activity of the ICM during the merger that is taking place along the
{\it SX2} axis. However, due to the very complex optical and X-ray
properties of A521, nothing can exclude that {\it A1} and {\it A2}
could be alternatively related to other merging events in A521, either
in its central field (e.g. the possible collisions of the clumps {\it
W} or {\it NE} with the main cluster), or in its outer regions not
covered by our \cha observations.

In this respect, several other substructures appear at larger scales
in the North of the iso-density map of the projected distribution of
the red sequence galaxies (see Fig.\ref{LSS}). New optical
observations (EFOSC2@3.6m ESO) have recently revealed the presence of
several galaxies at the cluster redshift in the two Northern clumps
{\it E1} and {\it E2} (Fig.\ref{LSS}), confirming that they are very
likely other merging subclusters at 1.5-2 \h Mpc from the cluster
centre. A radio relic in the South-East region of the cluster has also
been discovered through new VLA observations of A521 (Ferrari 2003,
see Appendix~\ref{radio} for more details).

\begin{figure*}
\centering
\resizebox{14cm}{!}  
{\includegraphics{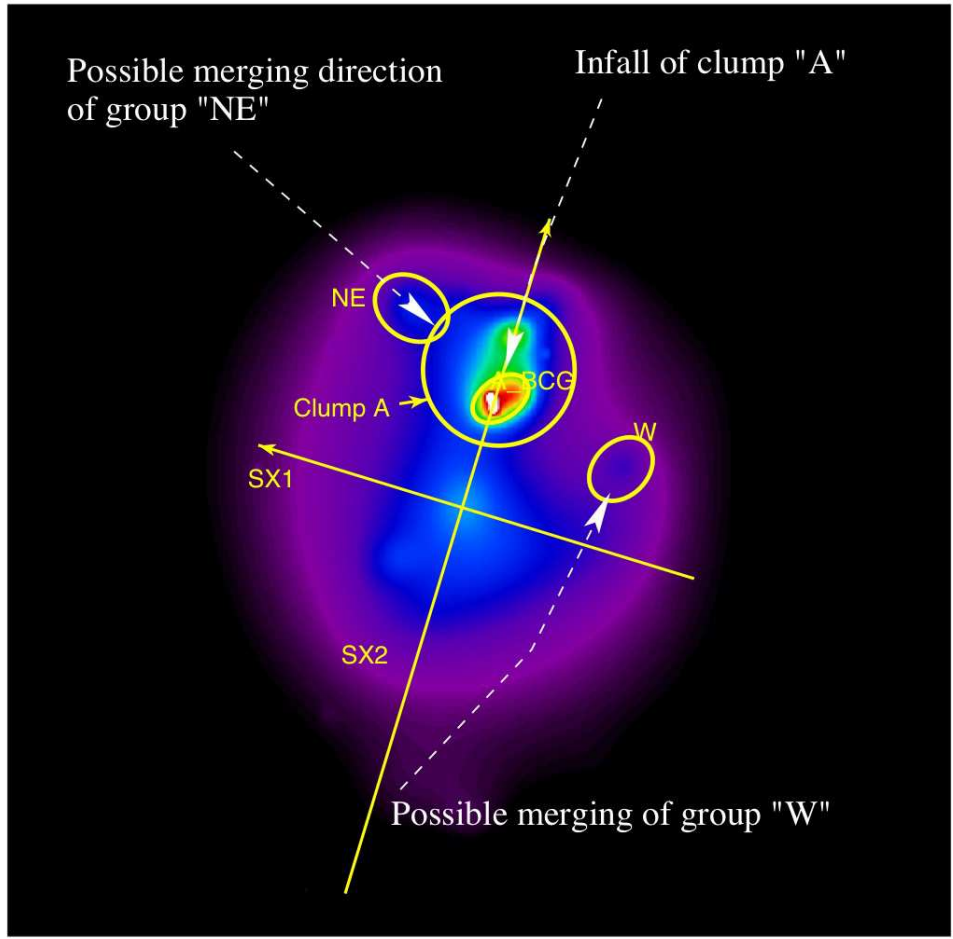}}  
\hfill  
\parbox[b]{18cm}{
 \caption{Cartoon showing the possible merging scenario in the central
 region of A521. The main merging subclusters (yellow) and their infall
 direction (white) are superimposed on the image of A521 X-ray diffuse
 emission.}
\label{cartoon}}  
\end{figure*}

\section{Summary and conclusions}\label{summ}

Through our {\it Chandra} observations of A521 we have confirmed that
this cluster is in a disturbed dynamical state, as shown by previous
X-ray and optical analysis (Arnaud et al. 2000; Maurogordato et
al. 2000; Ferrari et al. 2003). A sketch of the possible merging
scenario in the central field of the cluster covered by our {\it
Chandra} observations is shown in Fig.~\ref{cartoon}, in which the
following features emerge:

\begin{itemize}

\item [-] a main cluster centred on the X-ray/optical barycentre of the system;

\item [-] a group of galaxies (clump {\it A}) with its ICM density peak 
centred on the BCG, which is infalling on the main cluster along a
NW/SE direction ($\sim$-0.5 Gyr from the closest cores
encounter);

\item [-] two other structures possibly interacting with the main cluster 
({\it W} and {\it NE}), the former in the central phases of an
off-axis collision coming somewhere from S-SE, the latter at the
beginning of the interaction and coming from North-East. The nature of
these two substructures, and in particular of {\it NE}, is however
very uncertain;

\item [-] two edges in the ICM density ({\it A1} and {\it A2}), probably due 
to ongoing merging events either in the central field of the cluster
observed by {\it Chandra}, or in its outer regions.

\end{itemize}

In conclusion, {\it Chandra} observations confirm that A521 is made up
by several sub-clusters and groups of galaxies converging towards the
centre of the cluster and observed in different phases of their
merging process. The higher resolution density and temperature maps
allow to corroborate and refine the merging scenario of the group
hosting the BCG (i.e. clump {\it A}), and to identify new signatures
of other possible interactions (i.e. the groups {\it W} and {\it NE},
the arcs {\it A1} and {\it A2}). A deeper and wider optical
spectroscopic coverage is now necessary to understand the most
puzzling regions of this system and clarify its extremely complex
multiple merging scenario.

\appendix

\section{Radio emission in A521}\label{radio}

\begin{figure*}
\centering
\resizebox{18cm}{!} {\includegraphics{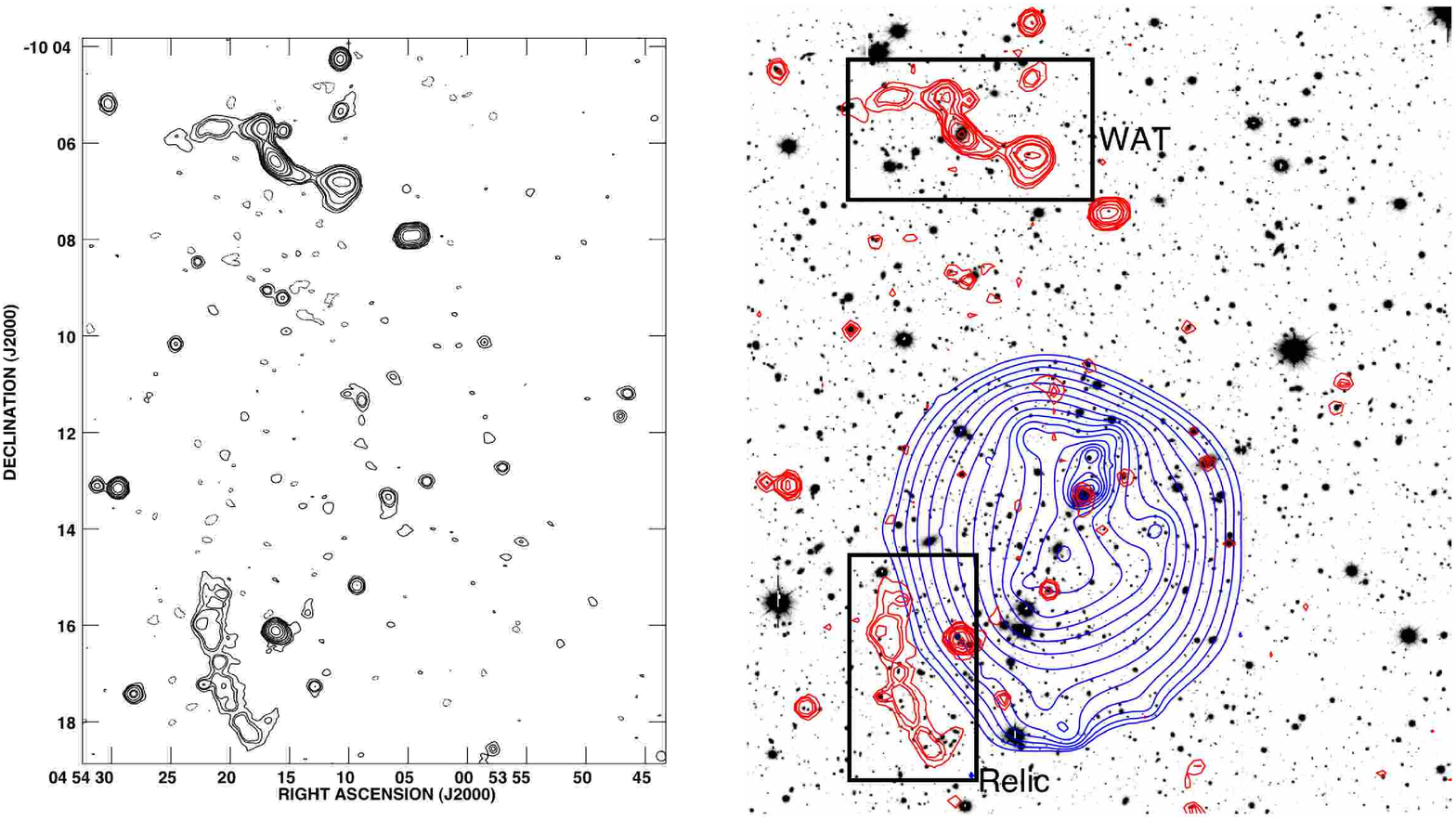}}
\hfill 
\parbox [b]{18cm}{
\caption{{\bf Left:} 20 cm radio map of A521 obtained from new VLA 
observations (Ferrari 2003). The contours levels are: -0.075, 0.075,
0.150, 0.200, 0.400, 0.800, 1, 2, 4, 8, 32 mJy/beam. The r.m.s. noise
level is 0.025 mJy/beam. The beam size is 12''$\times$12''. {\bf
Right:} contours of radio (red) and X-ray (blue) emission are
overlayed on the deep I-band image of A521 central field
($\sim15'\times13'$). The contour levels are set as in the left panel
(radio) and in Fig.~\ref{morf1} (X-ray). The newly detected radio
relic (Ferrari 2003) is the extended source in the South/East cluster
region. The WAT radio source in the North of the cluster is also
indicated. }
\label{fig:radio}} 
\end{figure*}

New VLA observations at 1.4 GHz, with angular resolution of
12''$\times$12'' and sensitivity of 0.025 mJy/beam (1$\sigma$) have
revealed the presence of a faint radio relic in the South/East region
of the cluster (Fig.~\ref{fig:radio} and Ferrari 2003), thus
supporting the perturbed dynamics of A521. Such low-brightness
extended radio sources are indeed only detected in cluster mergers
(Feretti 2003).  Detailed results on this radio source will be
presented elsewhere (Ferrari et al. in preparation).

A Wide Angle Tail (``WAT'') radio source has also been detected in the
North of the cluster (Fig.~\ref{fig:radio} and Ferrari 2003). The WAT
is located in the clump E1, which is probably merging with the main
cluster (see Sect.~\ref{other} and Fig.~\ref{LSS}). Further
spectroscopic observations could reveal if the optical galaxy
associated to the WAT is at the cluster redshift. The relative motion
between the host galaxy and the ICM, due to the infall of the clump E1
towards the cluster centre, would then be responsible for the observed
bend of the radio jets (Feretti and Venturi 2002).

\begin{acknowledgements}  
We warmly thank Wolfgang Kapferer, Magdalena Mair and Jean-Luc
Sauvageot for intensive and fruitful discussions on the merging
scenario of the cluster. We are very grateful to Luigina Feretti for
her helpful contribution to the analysis of the radio properties of
A521. The authors thank the anonymous referee for his/her suggestions
that improved the presentation of the paper. This research was
supported in part by Marie Curie individual fellowship
MEIF-CT-2003-900773 (CF).
\end{acknowledgements}

{}
 
\end{document}